\documentstyle[12pt]{article}

\def\lsim{\raise0.3ex\hbox{$\;<$\kern-0.75em\raise-1.1ex\hbox{$\sim\;$}}}
\def\gsim{\raise0.3ex\hbox{$\;>$\kern-0.75em\raise-1.1ex\hbox{$\sim\;$}}}

\def\u{\underline}

\begin{document}

\rightline{SISSA/EP/98/132}
\rightline{hep-ph/9812244}

\bigskip

\begin{center}

{\LARGE {\bf The Physics of Relic Neutrinos }}

\bigskip

{\large {\it Amol Dighe$^1$, Sergio Pastor$^2$, 
	Alexei Smirnov$^{1,3}$}}

\medskip

{$^1$ {\it The Abdus Salam International Center for
Theoretical Physics, \\ Strada Costiera 11,  34100, Trieste,
Italy.}}

\smallskip

{$^2${\it Scuola Internazionale Superiore di Studi Avanzati, \\
Via Beirut 4,  I-34013, Trieste, Italy.}}

\smallskip

{$^3${\it Institute for Nuclear Research, RAN,
Moscow, Russia.}}

\end{center}

\bigskip

\begin{abstract}
We report on the main results presented at the 
workshop on The Physics of Relic Neutrinos. 
The study of relic neutrinos involves
a broad 
spectrum of problems in particle physics, 
astrophysics and cosmology. 
Features of baryogenesis and leptogenesis  
could be imprinted in the properties of the relic neutrino sea. 
Relic neutrinos  played a crucial role in 
the big bang nucleosynthesis.
Being the hot component of the dark matter, 
they have participated in the structure formation in the
universe. Although the direct detection of the sea 
seems impossible at this stage, there could be  various 
indirect manifestations  of these neutrinos which 
would allow us to study  the properties of the sea both 
in the past and at the present epoch. 

\end{abstract}

\section{Introduction}

Neutrinos are 
one of the most abundant components
of the universe. 
Apart from the
3K black body electromagnetic radiation,
the universe is filled with a
sea of relic neutrinos which were created
in the early stages and decoupled from the rest of the matter
within the first few seconds.
These relic neutrinos have played a crucial role
in primordial nucleosynthesis, structure formation
and the evolution of the universe as a whole.
They  may also contain clues about
the mechanism of baryogenesis.

The properties of  relic neutrinos, their role in
nature and their possible manifestations were the main
topics of the workshop on The Physics of Relic Neutrinos.
It was organised at The Abdus Salam International Center for
Theoretical Physics (ICTP), Trieste, Italy during September
16 -- 19, 1998, by ICTP and INFN.

The workshop was attended by about 80 participants. Around 40
talks were distributed in the following sessions : 
$\bullet$ Neutrino Masses and Mixing
$\bullet$ Leptogenesis and Baryogenesis
$\bullet$ Big Bang Nucleosynthesis
$\bullet$ Structure Formation
$\bullet$ Detection and Manifestations of Relic Neutrinos
$\bullet$ Other sources of Neutrino Background
$\bullet$ Neutrinos in Extreme Conditions.

In what follows, we will describe the main results presented 
in the talks. We also give as complete as possible a list
of references to the original papers where the results have 
been published.

\section{Neutrino Masses and Mixing}

If neutrinos are massless and there is no significant lepton
asymmetry in the universe, the properties of the relic neutrino
sea are well known: the neutrinos are uniformly distributed in the
universe with a density of  113/cc/species and
at a temperature of 1.95K.

The existence of a nonzero neutrino mass
can dramatically change the properties of the
sea and its role in the evolution of the
universe. 
In this connection, 
the existing evidences for non-zero  neutrino masses and mixing
have been extensively discussed.

Recent SuperKamiokande (SK) results on
atmospheric neutrinos \cite{sk}
give the strongest evidence for a
nonzero neutrino mass.
\u{E. Lisi} (Bari) showed that the best fit to
the sub-GeV, multi-GeV and upward-going muon
data from SK is obtained at \cite{lisi}
$$
\Delta m_{23}^2 = 2.5 \times 10^{-3} \mbox{ eV}^2 \ , \
\sin^2 \theta_{23} = 0.63  \mbox{  and  } 
\sin^2 \theta_{e3} = 0.14~~,
$$
though taking the CHOOZ results into account will
decrease the values of $\sin^2 \theta_{e3}$
and $\Delta m_{23}^2$.
Maximal depth $\nu_\mu \leftrightarrow \nu_{sterile}$
oscillations can 
also give a good fit of
the data (\u{O. Peres}, Valencia), with a
slightly higher value of $\Delta m^2$ \cite{peres}.

A majority of alternative
explanations
of the atmospheric neutrino problem, like a neutrino decay,
reviewed by \u{S. Pakvasa} (Hawaii),
still imply nonzero neutrino masses.
The decay of
neutrinos can account for the sub-GeV and multi-GeV
atmospheric neutrino data rather well \cite{pakvasa}.
However in this case, the deficit of the
upward-going muon fluxes, as indicated by the data from
SK and MACRO, cannot be explained.

All the above explanations of 
the SK results imply that at least one neutrino 
species has the mass $m \geq 0.03$ eV.  
This  means that at least one of the components of the 
relic neutrino sea is non-relativistic, opening up the 
possibility of the structure formation of the sea.

Though the results on solar neutrinos 
give a strong hint of the existence of a nonzero neutrino mass,
we are still far from the final conclusion.
\u{L. Krauss} (Case Western) pointed out that if 
the oscillations are ``just-so'',
certain correlations between the spectral distortions and 
seasonal variations
of the solar neutrino signal may be observable
\cite{krauss}. 

If both the solar and the  
atmospheric neutrino anomalies have
the oscillation interpretation, 
neutrinos can contribute significantly to 
the hot dark matter only
if the neutrino mass spectrum is degenerate:
all three neutrinos have the mass of
about 1 eV. A strong bound on this scenario follows from the 
negative searches of the neutrinoless 
double beta decay. With a degenerate mass spectrum,
the present bound on the effective Majorana 
neutrino mass \cite{majorana} --
$$
m_{\rm majorana} < 0.45 \mbox{ eV  (90\% C.L.)} ~~
$$
-- implies a large mixing of the electron neutrinos 
and  some cancellation of contributions from different 
mass eigenstates. Alternatively it means that the 
neutrino contribution to 
the energy density in the universe is $\Omega_{\nu} < 0.06$.   
\u{F. Simkovic} (Comenius) showed that new estimations 
of the nuclear matrix elements
using the pn-RQRPA (proton -- neutron relativistic
quasiparticle random phase approximation)
allow the weakening of
the present bound on the Majorana neutrino masses by 50\%
\cite{simkovic}.

The reconstruction of the whole neutrino mass spectrum on the basis of
the present data is of a great importance both for particle physics and
cosmology. Several plausible patterns of neutrino masses and mixing 
have been elaborated. One possibility which has attracted significant
interest recently (especially in connection 
with the recent measurements of the recoil electron energy spectrum of
the 
solar neutrinos) is the bi-maximal mixing scheme with 
degenerate neutrinos \cite{vissani}. As was described by  
\u{F. Vissani} (DESY), this scheme reproduces 
$\nu_{\mu} \leftrightarrow \nu_{\tau}$ oscillation solution of the 
atmospheric neutrino problem, explains the solar neutrino data 
by ``just-so'' oscillations of $\nu_e$ into $\nu_{\mu}$ and
$\nu_{\tau}$, 
and gives a significant amount of the HDM without conflicting 
with the double beta decay bound.  
However, this scheme requires a strong fine-tuning.

\u{M. Fukugita} (Tokyo) reviewed the models of fermion masses 
based on the $S_{3 L}\times S_{3 R}$ permutation symmetry which 
lead to the ``democratic'' mass matrices for charged fermions.
The Majorana character of neutrinos 
admits a diagonal mass matrix 
with a small mass splitting due to the symmetry violation. 
In this case, one gets a large lepton mixing and neutrino mass degeneracy 
required for HDM. Fukugita 
presented the embedding of this scheme of 
mass matrix patterns in SU(5) GUTs
\cite{fukugita}.

\u{R. Mohapatra} (Maryland) showed that the bi-maximal mixing 
pattern can be derived from the maximal, symmetric,
four-neutrino mixing in the limit that one of the neutrinos is
made heavy \cite{rabi1}.  
He also showed that combining the permutation
symmetry $S_3$ with a $Z_4 \times Z_3 \times Z_2$
symmetry in the left-right symmetric extension
of the standard model, the mixing pattern of
the democratic mass matrix can be generated \cite{rabi2}.
This would account for a large 
$\nu_\mu \leftrightarrow \nu_\tau$
and maximal $\nu_e \leftrightarrow \nu_\mu$ mixing, along with
small Majorana masses through the double seesaw mechanism.

The attempts to accomodate all the existing data
and / or to explain the large lepton mixing  lead
to the introduction of sterile neutrinos (\u{Mohapatra})
\cite{rabi-zurab}.
Their existence would have enormous implications
for astrophysics and cosmology. 
\u{Z. Berezhiani} (Ferrara) explored the possibility of
the sterile neutrinos $\nu'$ being from a mirror
world  
which communicate with our 
world  only through gravity or 
through the exchange of some particles of 
the Planck scale mass 
\cite{zurab}. 
In the mirror world, the 
scale of the electroweak symmetry breaking can be higher  
than  our scale: $v_{EW}' = z \cdot v_{EW}$, $z > 1$. 
In this case   
$\nu_e \leftrightarrow \nu'_e$ mixing  can
provide the solution of the solar neutrino problem  via the MSW 
effect  ($z \sim 30$) or ``just-so'' oscillations ($z \sim 1$). The mirror 
neutrinos (and mirror baryons) can also form the
dark matter in the universe. 

Various aspects of the theory of neutrino oscillations, 
and in particular the problem   of coherence and
decoherence in the oscillations,
were discussed by \u{L. Stodolsky} (MPI, Munich).

\section{Leptogenesis and Baryogenesis}

In the early universe, one of the first processes directly
influenced by the neutrinos would have been
those of leptogenesis 
and baryogenesis.
One of the favoured mechanisms 
for the dynamical generation of the 
observed baryon asymmetry  
is through the production of a lepton asymmetry,
which can then be converted to
the baryon asymmetry by $(B-L)$ conserving
electroweak
sphalerons \cite{review}.

The leptonic asymmetry can be generated
in the CP-violating decay of heavy
($M \gsim 10^{10}$ GeV)
right handed neutrinos $N_i$ to Higgs and usual 
neutrinos $N_i \to \ell^c H^*$, $\ell H \to N_i$. 
The lifetime of these 
Majorana neutrinos needs to be
long enough, so that the thermal equilibrium is broken. 
\u{E. Roulet} (La Plata) discussed the
finite temperature effects on
these CP violating asymmetries \cite{roulet}.

\u{E. Akhmedov} (ICTP) described a new  scenario 
of baryogenesis via neutrino oscillations \cite{akhmedov}.  
The lepton asymmetry  is created in
CP-violating oscillations of 
three right handed neutrino 
species with masses 20 -- 50 GeV. 
The neutrinos should have 
very small ($10^{-8} -  10^{-7}$) and different 
Yukawa couplings. 
These Yukawa couplings lead both
to the production of the RH neutrinos 
and  the propagation of  the generated asymmetry to the 
usual leptons. 
The lepton asymmetry 
is generated in different neutrino species,
but the total lepton number is still zero. 
At least one of the
singlet neutrino species needs to be in equilibrium 
and at least one out of equilibrium when the
sphalerons freeze out. Then only those 
neutrinos which are in equilibrium will transform the asymmetry 
to light ($SU(2)$ doublet) leptons. This asymmetry will be then 
converted to the baryon asymmetry. 
Thus, a lepton asymmetry can be produced 
without a total lepton number
violation, through the ``separation'' 
of charges.

\u{A. Pilaftsis} (MPI, Munich) talked about a model with
two singlet  neutrinos per fermion family,
which get their masses through an off-diagonal
Majorana mass term \cite{pilaftsis}. The mass
splitting between these two neutrinos can be small  
(in $E_6$ theories, for example) -- as small as
10 -- 100 eV for the Majorana
neutrino masses of 10 TeV. If
the splitting  is comparable to the
decay widths, CP asymmetries in the neutrino decays are
resonantly enhanced. A remarkable consequence is that
the scale of leptogenesis may be lowered upto the TeV range.

In all the above scenarios, the seesaw mechanism 
leads to the light neutrino masses 
in the range $10^{-3} - 1$ eV,
which  are
relevant for cosmology and
for explaining the solar and atmospheric neutrino data.

The leptonic asymmetry can also be produced without
right handed neutrinos in the decays of 
two heavy Higgs triplets 
(\u{U. Sarkar}, PRL, Ahmedabad). The Higgs masses 
of $10^{13}$ GeV 
lead both to a successful leptogenesis and to 
a few eV scale for masses of the
usual neutrinos \cite{utpal}.

In all the above scenarios, the leptonic asymmetry is
of the same order as the final baryon asymmetry. 
In general, a large lepton asymmetry can be produced
without a large baryon asymmetry, 
e.g. through the Affleck-Dine mechanism \cite{affleck}.

In the scenario with the decay of the RH neutrinos,
if the hierarchy of the Dirac masses as well as
the Majorana masses is similar to
that of the up-type quarks, and if the solar neutrino
deficit is due to the MSW effect,
the temperature for baryogenesis 
may be as high as $T_B \sim M_R \sim 10^{10}$ GeV.
At such high temperatures, however,  
a large number of gravitinos are 
generated. 
These gravitinos 
might overclose the universe, 
and if they decay late, 
modify the primordial light element abundances
in a way that is incompatible with observations. 
According to \u{W. Buchm\"uller} (DESY), these problems can be avoided  
if gravitinos are the LSPs (and therefore stable), and
have the mass 10 -- 100 GeV.
The relic density of these gravitinos 
will be cosmologically important
and they can play the role of the cold dark matter \cite{buchmuller}.

\section{Big Bang Nucleosynthesis}

Properties of the neutrino sea are crucial for the outcome 
of the big bang nucleosynthesis (BBN), i.e. the primordial
abundances of the light nuclides: 
D, $^3$He, $^4$He and $^7$Li.

The implications of the recent data on
the primordial abundances
for cosmology and particle physics were reviewed by 
\u{G. Steigman} (Ohio)
\cite{Steigman,revBBN}. The data appear to be in rough agreement
with the predictions of the standard cosmological model for three
species of light neutrinos and a nucleon-to-photon ratio 
restricted to a narrow range of 
$$
\eta \equiv n_B/n_\gamma =  (3 - 4) \times 10^{-10}.
$$
A closer inspection, however, reveals a tension between the inferred
primordial abundances of D and $^4$He.  For deuterium, at
present there are 
two different analyses of the data from the observations of
high-redshift, low-metallicity absorbing regions: the first analysis
leads to the primordial abundance of  \cite{hiD}
$$
D/H = (1.9 \pm 0.5) \times 10^{-4} \; (\mbox{ high } D),  
$$
while the second gives \cite{lowD}
$$
D/H  = (3.40 \pm 0.25) \times 10^{-5} \;  (\mbox{ low } D).
$$  
The primordial abundance of
$^4$He is derived from the observations of low-metallicity extragalatic H
{\sc II} regions. Here also, there are two inconsistent results
for the $^4$He mass abundance $Y_P$. One of the calculations
\cite{hiHe}  leads to a high number 
$$
Y_P = 0.244 \pm 0.002 \; (\mbox{ high }^4 \mbox{He}) ,
$$ 
and another \cite{lowHe} gives a low number,
$$
Y_P = 0.234 \pm 0.002 \; (\mbox{ low }^4 \mbox{He}). 
$$

The consistency of the D and $^4$He results with the predicted
abundances in the standard BBN is possible in two cases:
(i) low D, high $^4$He and high $\eta$,  or
(ii) high D, low $^4$He and low $\eta$.

Resolution of this conflict may lie within the statistical
uncertainties in the data or with the systematic uncertainties: 
in the extrapolation from 
``here and now to there and then". 
However, if both the D and $^4$He
abundances are   low, the Standard BBN is in ``crisis''. The 
problem can be resolved if the contribution of some
non-standard particle
physics leads to an effective number of light neutrino species 
($N^{\nu}_{\rm eff}$) at the
time of BBN smaller than three. This can be realized, for example, if
the mass of the tau neutrino is in the range of a few MeV and it
decays invisibly with $\tau \lsim 5$ sec (\u{S. Pastor}, Valencia). 
In fact, $N^{\nu}_{\rm eff}$ can be as low as 1
if the products of the neutrino decay include electron neutrinos, due to
their direct influence on the 
neutron $\leftrightarrow$ proton reactions \cite{Pastor}.

A simple statistical method for determining the correlated
uncertainties of the light element abundances expected from BBN was
presented by \u{F. Villante} (Ferrara) \cite{Villante}. This method,
based on the linear error propagation, avoids the need for lengthy Monte
Carlo simulations and helps to clarify the role of the different
nuclear reactions. The results of a detailed calculation of nucleon
weak interactions relevant for the neutron-to-proton ratio at the
onset of BBN were presented by \u{G. Mangano} (Naples) \cite{Mangano}.

The presence of sterile neutrinos in the relic neutrino sea can
significantly modify BBN. 
Though recent conservative 
bounds on $N^{\nu}_{\rm eff}$ still admit 
more than four neutrino species \cite{olive},
the question  of whether sterile neutrinos 
can be in equilibrium
at the BBN is still alive.

If sterile neutrinos have masses and mixing which 
give the solution of the atmospheric neutrino anomaly, then 
the equilibrium concentration of sterile neutrinos 
will be generated via
$\nu_{\mu} \leftrightarrow \nu_{s}$ oscillations. 
This can be avoided if the lepton asymmetry 
of the order $\gsim 10^{-5}$ exists 
at the time of 
$\nu_{\mu} \leftrightarrow \nu_{s}$ oscillations \cite{fvprl95}. 
The asymmetry can be produced in the oscillations 
$\nu_{\tau} \leftrightarrow \nu_s$ and 
$\overline{\nu}_{\tau} \rightarrow
\overline{\nu}_s$ at earlier times. 
The numerical integrations of the corresponding quantum
kinetic equations (\u{R. Volkas}, Melbourne)
show that this requires $m_{\nu_\tau} \gsim 4$ eV
(for $|\delta m_{atm}^2|=10^{-2.5}$ eV$^2$)
\cite{Volkas3}. However \u{X. Shi} (San Diego) concludes from his
calculations that a $\nu_\tau$ with a larger mass, $15 \mbox{ eV }
\lsim m_{\nu_\tau} \lsim 100 \mbox{ eV }$, is neeeded. Such a $\nu_\tau$ 
must decay non-radiatively with a lifetime $\lsim 10^3$ years, in
order to have a successful structure formation at high redshifts
\cite{ShiBBN}.  
Recently Shi's results have been criticized by Foot
and Volkas \cite{Volkas4}, who 
confirmed their previous lower value of $m_{\nu_\tau}$.

Volkas also presented the general principles of the creation 
of  a lepton 
asymmetry as a generic outcome of active to sterile neutrino
oscillations ($\nu_a \rightarrow \nu_s$ and $\overline{\nu}_a 
\rightarrow \overline{\nu}_s$, where $a=e,\mu,\tau$) in the
early universe as a medium. It can be studied from a 
simpler, Pauli-Boltzmann approach as well as
starting from the exact quantum kinetic equations \cite{Volkas1}.  
If a significant 
electron-neutrino asymmetry ($\gsim 1\%$) is generated,
$N^{\nu}_{\rm eff} $
can be less than three \cite{Volkas2}.

\u{D. Kirilova} (Sofia) discussed the oscillations 
of $\nu_a \leftrightarrow \nu_s$ with a small 
mass difference  ($\delta m^2 < 10^{-7}$ eV$^2$).
These oscillations 
become effective after the decoupling of active
neutrinos.  Using an exact kinetic approach, it is possible to study
the evolution of the neutrino number density for each momentum mode.
This approach allows one to calculate all the effects of neutrino
oscillations on the production of primordial $^4$He: the depletion
of the neutrino population, the distortion of the energy spectrum and
the generation of a neutrino asymmetry \cite{Kirilova}.

\section{Structure Formation}

Neutrinos are a major component
of the hot dark matter (HDM) -- the particles
which were relativistic at 
$t \sim 1$ year, when $T \lsim$ keV and
the ``galaxies'' came within the horizon. 
The neutrinos with masses in the eV range would
contribute significantly to the matter density 
in the universe:
$$
\Omega_\nu = 0.01 ~ h^{-2} \left( \frac{m_{\nu}}{\rm eV} \right)~~, 
$$
and even smaller masses can be
relevant for the structure formation. 
For $\Omega_\nu \ge 0.1$, neutrinos would 
significantly influence the observable spectrum of density 
perturbations, giving more strength to 
supercluster scales and suppressing smaller scales.

The primordial density fluctuations 
in the universe are
probed, in particular, by the anisotropies in the 
cosmic microwave background (CMB) radiation 
(for scales $\gsim$ 100 Mpc), and
observations of the large scale distribution of galaxies.
Optical red shift surveys of galaxies can now examine
scales upto $\sim$ 100 Mpc.
As described by \u{J. Silk} (Berkeley),
no current model seems to fit the detailed shape
of the power spectrum of the primordial
density perturbations 
and satisfy all the existing constraints,
although the Cold + Hot dark matter (CHDM) model
with  
$$
\Omega_{cold} \sim 0.7 ,~~~~ \Omega_\nu \sim 0.2,~~~~~
\Omega_{b} \sim 0.1 
$$
gives a relatively better fit \cite{silk-sci}.
This model implies a neutrino mass (or the
sum of the neutrino masses) of about
5 eV and describes the nearby universe well, however
it (like the other models with $\Omega = 1$ and zero
cosmological constant $\Lambda$)
is disfavoured by the new data on 
(i) the early galaxies, 
(ii) cluster evolution, and 
(iii) high redshift type IA supernovae.

The models with a cosmological constant,
$\Lambda$CDM ($\Omega_\Lambda \approx 0.6$),
seems to be favoured in the light of the new data
\cite{primack1},
but the overall fit is still not satisfactory.

The sizes of voids give
an important clue for the relative
fraction of the HDM.
\u{J. Primack} (UC Santa Cruz) 
described the use of
the void probability function (VPF) 
to quantify this distribution \cite{vpf}. 
It is found that on intermediate (2 -- 8 $h^{-1}$ Mpc)
scales, the VPF for the standard CHDM model 
(with $\Omega_{cold} / \Omega_{hot} / \Omega_{bar} = 
0.6 / 0.3 / 0.1 $)
exceeds the observational VPF,
indicating that the HDM fraction is lower than what was thought
earlier.

\u{T. Kahniashvili} (Tbilisi) argued that that consistency with the
current data can be achieved for the (COBE-normalized)
models only for 
$$
\Omega_{hot} /  \Omega_{matter} \le 0.2,
\  h = 0.5 (0.7),  \mbox{ and } 0.45 (0.3) \le \Omega_{matter}
\le 0.75 (0.5) 
$$ 
at $1 \sigma$ level \cite{tina},
so that $\Omega_{\nu} < 0.1$. 

The presence of a non-zero cosmological constant, 
though theoretically problematic from the point
of view of ``naturalness'', 
seems to
help in understanding the large scale structure better.
In that case, the main conclusion 
(as emphasized by \u{M. Roos}, Helsinki) is that 
the presence of HDM is
no longer necessary (and eV neutrinos are not needed to
provide this component), although some amount of HDM
is still possible and may be
useful for a further tuning.

The situation can be clarified with new precision measurements
of the CMB
anisotropy by MAP and PLANCK, which
will be sensitive to $\Omega_\nu \sim 0.01$ and therefore
$m_\nu \gsim 0.2$ eV \cite{cmbr}.
\u{S. Hannestad} (Aarhus) discussed the role of these  
in constraining neutrino decays and
for detecting the imprints of sterile neutrinos
\cite{hannestad}. PLANCK will be able to probe the
anisotropy to the multipole $l \lsim 2500$, so that the number of
neutrino species can be determined to a precision of
$\Delta N_{\nu} \sim 0.05$, which is much better than
that obtained from the BBN.  
New galaxy surveys like SDSS will probe 
neutrino masses as low as 0.1 eV \cite{tegmark}.

The CMB anisotropy measurements also allow us to
put a limit on the degeneracy of neutrinos.
According to \u{S. Sarkar} (Oxford),  
the present CMB data 
still admits a rather 
strong degeneracy
(the best fit being at 
$\mu/T = 3.4$ with the spectral index $n = 0.9$), 
and hence a large lepton asymmetry.
The existence
of such a large lepton asymmetry can modify the history
of the Universe, leading to  symmetry 
non-restoration at high temperature and thus solving
the monopole and domain wall problems \cite{bajc}.
The height of the lowest multipole peak in the CMB spectrum
increases with the degeneracy of neutrinos.
(The difference in heights between the $\mu/T = 1$
and  $\mu/T = 0$  cases
is about 10\%.)
So forthcoming precision measurements
of the multipole spectrum will be able
to restrict the degeneracy.

\section{Detection and  Manifestations of Relic Neutrinos}

The direct detection of relic neutrinos will of course be of
a fundamental importance. However it looks practically 
impossible with the present methods. 
The situation have been summarized  
several years ago in the review \cite{smith}, 
and some possible schemes have been proposed in \cite{speake}.
At the same time, it is possible to search for 
some indirect manifestations of
the relic sea even now.

\u{D. Fargion} (Rome) \cite{Fargion} and 
\u{T. Weiler} (Vanderbilt)
\cite{Weiler} have considered a mechanism 
involving relic neutrinos that may generate the
highest energy cosmic rays detected at the earth 
(see for example \cite{AGASA}), which have energies 
above the Greisen-Zatsepin-Kuzmin (GZK) cut-off of
$\sim 5\times 10^{19}$ eV \cite{GZK}.  The process is the annihilation
of ultrahigh energy neutrinos on  the
nonrelativistic neutrinos from the relic sea:
$$
\nu_{\mbox{cosmic}}+ \bar{\nu}_{\mbox{relic}} \rightarrow Z
\rightarrow \mbox{nucleons and photons .}
$$
For the neutrino mass $m_\nu \sim$ few eV, the energy 
of cosmic ray neutrinos should be about $E_\nu \gsim 10^{21}$ eV. 
It is assumed that the production rate is greatly enhanced 
due to a significant
clustering of the relic neutrino density in the halo of our galaxy or
the galaxy cluster.  The secondary  nucleons and photons 
may propagate to the earth without too much energy
attenuation and are the  primary candidate particles for inducing
super-GZK air showers in the earth's atmosphere. A numerical
calculation has been done in  \cite{Yoshida}
which indicates that such cascades could contribute more than 
10\% to the observed cosmic ray flux above $10^{19}$ eV  
in the case of eV neutrinos.
Recently Waxman \cite{waxman}
has showed that 
for the annihilation to contribute significantly to the detected 
cosmic ray-events, a new class of high energy neutrino
sources, unrelated to the sources of UHE cosmic rays,
needs to be invoked.

The relic sea could also be detected if neutrinos are massive and
undergo a radiative decay. This hypothesis 
was suggested to explain the high ionization of the 
interstellar hydrogen. 
Present status of this hypothesis was summarized by \u{D. Sciama}
(Trieste) \cite{Sciama}.
If this heavy (27.4 eV) neutrino is sterile, it will 
decouple earlier (at $T \ge 200$ MeV) and its contribution
to the matter density $\Omega$ will be small,
thus avoiding any conflict with the structure formation
\cite{rabi-sciama}.
Direct searches of the expected EM line at $\lambda \sim 900$ \AA~from
this radiative decay are being performed by 
EURD detector 
and the results are expected soon. 
The decaying neutrino cosmology leaves a particular imprint in the
angular power spectrum of temperature fluctuations in the CMB, which
will be tested with the forthcoming MAP and PLANCK surveyor missions.

The evolution of the relic neutrino sea, the possibility of 
clustering, the formation of structures, 
local concentrations etc.  are of
great 
importance both for direct and indirect detections of relic neutrinos.
A possible scenario of the structure 
formation on galactic scales was discussed by \u{N. Bilic}
(Zagreb):
self-gravitating neutrino clouds can show
``gravitational phase transitions'' in the process of contraction
and form neutrino stars, the scale of whose sizes
would depend on the neutrino mass \cite{bilic}.

\section{Other Sources of Neutrino Background}

Apart from the big bang relic neutrinos, the present universe is 
filled with relic neutrinos from astrophysical sources: past
supernovae, supermassive objects and
probably, primordial black holes.

The possibilities of the detection of neutrinos from relic and
real-time supernovae with existing and new detectors were discussed by
\u{D. Cline} (UCLA) and \u{K. Sato} (Tokyo).  
Sato has calculated the expected
rate of relic supernova
neutrinos at the Super-Kamiokande detector.
The rate of supernova explosions is
derived from a model of galaxy evolution where the effect of 
the chemical
evolution is appropriately taken into account \cite{Sato}.
Monte-Carlo simulations show that the rate is a few events/year in the
observable energy range of 15-40 MeV, which is still 
about two orders of magnitude smaller
than the observational limit at Super-Kamiokande. A similar rate is
found for the new experiment ICARUS, described by Cline.

A future detection of a supernova neutrino burst by large underground
detectors 
will provide a measurement of neutrino masses and mixing 
(Cline).  New projects of a
supernova burst observatory (SNBO/OMNIS) with an
operation time of $\gsim 20-40$
years were  described, where neutrinos will be detected through the 
secondary neutrons emitted by the recoiling nuclei \cite{Cline}.

A new cosmic neutrino source may be provided by Supermassive Objects
(SMOs), that may be formed as the final evolutionary stage of 
dense star clusters (\u{X. Shi}, San Diego). Through relativistic
instabilities, SMOs will eventually 
collapse into giant black holes, such
as those at the centers of galaxies.
A significant fraction of the gravitational binding energy of the
collapse of the SMOs may be released by freely escaping neutrinos in a
short period of time ($\sim 1$ sec) with an average energy 1-10 MeV.
Neutrino bursts from nearby SMO's ($d \leq 750$ Mpc) may be detectable
at ICECUBE, a planned $1$ km$^3$ neutrino detector in Antarctica (an
expanded version of the current AMANDA) with an expected
 rate of $\sim 0.1$ to
$1$ burst per year \cite{ShiSMO}.

Some contribution to the relic neutrino sea may also come from the
evaporation of Primordial Black Holes (PBHs)
through Hawking radiation \cite{Bugaev}. 
\u{E. Bugaev} (Moscow) showed that the most favorable
energy to detect the 
flux of neutrinos of PBH origin is
a few MeV. Comparison of the theoretically expected
neutrino flux from PBHs with Super-Kamiokande data 
sets an upper bound on the contribution of
PBHs to the present energy density of the universe
($\Omega_{PBH} \lsim 10^{-5}$). This,  however, is  much weaker than the
bounds from the $\gamma$-background data.

\section{Neutrinos in Extreme Conditions}

An important aspect of the physics of the relic neutrinos is 
the propagation and the interactions of neutrinos 
in the extreme conditions of
the very hot and dense plasma, in 
strong magnetic fields, etc..

\u{R. Horvat} (Zagreb) has used the real-time approach of the thermal
field theory (TFT) to calculate
the finite temperature and finite
density radiative corrections to the neutrino effective 
potential in the CP-symmetric early universe
(see also \cite{horvat}). 
The $\cal{O}(\alpha)$ photon corrections 
have been shown 
to be free of infrared and finite mass singularities,
so that bare purturbation series is 
adequate for the calculations.

\u{D. Grasso} (Valencia) 
 has calculated the radiative decay rate of 
neutrinos in a medium using a generalisation of 
the optical theorem \cite{Grasso}. This is a
powerful method to handle dispersive and dissipative properties of the
medium. The results are applicable to the neutrino
evolution in the early universe where the electron -- positron
plasma is ultra-relativistic and non-degenerate.

\u{A. Ioannisian} (Munich) discussed the \v{C}erenkov radiation process
$\nu \to \nu \gamma$ in the presence of a homogeneous
strong magnetic field \cite{ara}. 
Apart from inducing an effective
neutrino-photon vertex, the magnetic field also modifies the photon
dispersion relations. Even for fields as large as
$B_{\rm crit} \equiv m_e^2/e \approx 4 \times 10^{13}$ Gauss
(which are encountered around pulsars), the \v{C}erenkov 
rate is found to be small, which indicates that the magnetosphere 
of a  pulsar is quite transparent to neutrinos.

\section{Summary and Outlook}

\quad 1. 
The SK atmospheric neutrino results
imply that neutrinos are massive and 
at least one component of the relic sea is
non-relativistic. This opens up the possibility 
of clustering of neutrinos and the formation of structures. 

Forthcoming experiments on atmospheric and 
solar neutrinos, double beta decay, etc. 
may shed more light on the neutrino mass spectrum, 
and therefore, on the relevance of neutrinos for 
cosmology.

The possible discovery of sterile neutrinos 
(light singlet fermions) that mix with 
usual neutrinos 
will have an enormous impact on astrophysics
and cosmology.

\medskip

2. The simple mechanism 
of the baryon asymmetry generation  via leptogenesis  
seems very plausible. 
Moreover, in several suggested scenarios, 
the masses of light neutrinos are expected to be in the
range relevant for cosmology. 

Further developments in this field would be related 
to the identification of the mechanism of 
neutrino mass generation as well as the studies of alternative
scenarios of baryogenesis -- like the electroweak baryogenesis 
based on supersymmetry.

\medskip

3. The neutrino sea 
has a strong influence on the big bang
nucleosynthesis.
Here the observational situation is not clear.
Conservative bounds admit more than 
four neutrino species
in equilibrium at the time of BBN, so that one light 
sterile neutrino in equilibrium is possible.
On the other hand, if the observations imply a
lesser number of effective neutrino species,
it can be accounted for by 
scenarios like neutrino decay or oscillations 
into sterile components. 

The progress would come from further studies of the
systematics in the determination of abundances, 
restrictions on $\eta$, and searches for sterile
neutrino effects in the laboratory experiments.

\medskip   

4.
 Recent cosmological data is changing 
our understanding of the 
role of neutrinos as the HDM: 
it seems that the HDM is not necessary, although 
some amount is allowed and may be useful for 
a better fit to the data. 

Future cosmological observations will  give
important information 
about the neutrino masses, the presence of sterile states, 
neutrino degeneracy, etc.. 

\medskip

5.
 The direct detection of the relic neutrinos is 
a challenge.
However, indirect observations of the 
neutrino sea are possible via the studies of the 
cosmic rays of ultrahigh energies, or through
the searches for radiative decays of relic neutrinos.

\medskip

There are deep connections between the physics of
relic neutrinos and a variety of fundamental open
questions in cosmology, astrophysics and particle physics.
Understanding the properties of the relic neutrino sea
and its possible detection will be one of the challenges for
the physics and astrophysics of the next millenium.

\section{Epilogue}

This report is an attempt to substitute the 
``Proceedings'', which are, in many cases, 
a nightmare for the organisers, a waste of time for
the speakers and a practically useless showpiece for the 
readers due to the time delays. 
Its objectives were 
\begin{itemize}
\item to give general information about the meeting 
(format, participants,  topics, etc.),
 
\item to review the  results and discussions,

\item to give, as much as possible, a complete 
reference list to the original papers 
of participants in which the 
results  presented during the conference 
were published. (Indeed, a majority 
of the results have been published 
before or within about two months after the meeting.) 
We also  give some information about 
other appropriate papers, as well as about
further related  developments during the short time
after the conference. 

\end{itemize}

This review has been written (as an experiment) by 
the organisers of the workshop. 
Probably a better idea would be to select ``reporters'' from among
the participants in advance, who will review the conference
in a short period of time.

\end{document}